\documentstyle[twoside,titlepage,psfig,11pt]{article}
\makeatletter
\def\section{\@startsection {section}{1}{\z@}{-3.5ex plus -1ex minus
    -.2ex}{2.3ex plus .2ex}{\large\bf}}
\def\subsection{\@startsection{subsection}{2}{\z@}{-3.25ex plus -1ex minus 
   -.2ex}{1.5ex plus .2ex}{\normalsize\bf}}
\def\subsubsection{\@startsection{subsubsection}{3}{\z@}{-3.25ex plus
   -1ex minus -.2ex}{1.5ex plus .2ex}{\normalsize}}
\makeatother
\begin{document}
\font\utap=cmmib10 scaled\magstep2
\font\urc=cmmib10 scaled\magstephalf
\font\smbf=cmbx10 scaled\magstep2
\font\smit=cmti12 
\font\fontname=cmcsc10 
\newcommand{\gsim}{\mbox{\raisebox{-1.0ex}{$\stackrel{\textstyle >}
{\textstyle \sim}$ }}}
\newcommand{\lsim}{\mbox{\raisebox{-1.0ex}{$\stackrel{\textstyle <}
{\textstyle \sim}$ }}}
\newcommand{\gtsima}{$\; \buildrel > \over \sim \;$}
\newcommand{\ltsima}{$\; \buildrel < \over \sim \;$}
\newcommand{\simgt}{\lower.5ex\hbox{\gtsima}}
\newcommand{\simlt}{\lower.5ex\hbox{\ltsima}}
\newcommand{\himpc}{{\hbox {$h^{-1}$}{\rm Mpc}} }
\newcommand{\bfq}{{\mbox{\boldmath $q$}}}
\newcommand{\bfr}{{\mbox{\boldmath $r$}}}
\newcommand{\bfx}{{\mbox{\boldmath $x$}}}
\newcommand{\bfy}{{\mbox{\boldmath $y$}}}
\newcommand{\bfv}{{\mbox{\boldmath $v$}}}
\newcommand{\bfpsi}{{\mbox{\boldmath $\psi$}}}
\newcommand{\bfPsi}{{\mbox{\boldmath $\Psi$}}}
\newcommand{\mPsi}{{\mit\Psi}}
\def\pp{\par\parshape 2 0truecm 15.5truecm 1truecm 14.5truecm\noindent}
\renewcommand{\theequation}{\mbox{\rm 
{\arabic{section}.\arabic{equation}}}}
\begin{titlepage}
\vspace*{-0.6cm}
\hspace*{12.5cm}RESCEU-26/97

\hspace*{12.5cm}OCHA-PP-100

\hspace *{12.5cm}UTAP-265/97


\hspace *{12.5cm}\today

\vspace*{1cm}
\baselineskip=14pt
\begin{center}
  {\large \bf  ACCURACY OF NONLINEAR APPROXIMATIONS IN SPHEROIDAL
  COLLAPSE  \\[5mm]
--- Why are Zel'dovich-type approximations so good? --- }

\vspace{1cm}
Ayako {\sc Yoshisato}${}^{1}$, Takahiko {\sc
    Matsubara}${}^{2,3}$ and Masahiro {\sc Morikawa}${}^{1}$
\end{center}

\vspace{0.2cm}
\noindent ${}^{1}${\it Department of Physics, Ochanomizu University,
2-2-1, Otsuka, Bunkyo-ku 112, Japan.}

\vspace{0.2cm}
\noindent ${}^{2}${\it Department of Physics, The University of Tokyo,
Tokyo 113, Japan.}

\vspace{0.2cm}
\noindent ${}^{2}${\it Research Center for the Early Universe, Faculty
of Science, The University of Tokyo, Tokyo 113, Japan.}
\vspace{0.3cm}

\vspace{0.5cm}

\centerline {\bf ABSTRACT}
\baselineskip=12pt
Among various analytic approximations for the growth of density
fluctuations in the expanding Universe, Zel'dovich approximation and
its extensions in Lagrangian scheme are known to be accurate even in
mildly non-linear regime.  The aim of this paper is to investigate the
reason why these Zel'dovich-type approximations work accurately beyond
the linear regime from the following two points of view: (1)
Dimensionality of the system and (2) the Lagrangian scheme on which
the Zel'dovich approximation is grounded.  In order to examine the
dimensionality, we introduce a model with spheroidal mass
distribution.  In order to examine the Lagrangian scheme, we introduce
the Pad\'e approximation in Eulerian scheme.  We clarify which of
these aspects supports the unusual accuracy of the Zel'dovich-type
approximations.  We also give an implication for more accurate
approximation method beyond the Zel'dovich-type approximations.

\medskip
\pp {\bf Subject headings}: cosmology: theory --- galaxies: clustering --- gravitation
--- large-scale structure of universe
\end{titlepage}

\renewcommand{\thefootnote}{\fnsymbol{footnote}}
\setcounter{footnote}{1}

\baselineskip=18pt
\setcounter{page}{2}

\section{Introduction}
\setcounter{equation}{0}

As an analytic approximation for the growth of density fluctuations in
the expanding Universe, Zel'dovich approximation (ZA hereafter) is
known to be unusually accurate even in mildly non-linear regime for
unknown reason (Zel'dovich 1970; 1973).  Recently the extensions of
this ZA (Post- and Post-post- Zel'dovich Approximations, PZA and PPZA
hereafter) have been developed (e.g., Bernardeau 1994; Buchert 1994;
Bouchet et al. 1995).  These Zel'dovich-type approximations are
confirmed to be far better than any other analytic approximations in
Munshi, Sahni \& Starobinsky (1994), Sahni \& Shandarin (1995) and
Sahni \& Coles (1996) using a spherical-collapse model.  However so
far, there has been no clear explanation why these Zel'dovich-type
approximations are so accurate beyond their range of applicability.
In this paper, we would like to address this issue from limited
aspects and with limited models.

In the Zel'dovich-type approximations, we work on the Lagrangian
coordinate scheme in which the location of a mass element ${\bf x}$ of
the fluid is expressed by the initial location ${\bf q}$ and the time
dependent displacement vector $\bfPsi$ as
\begin{equation}
   \bfx(\bfq,t) = \bfq + \bfPsi(\bfq,t).
   \label{defpsi}
\end{equation}
Then the density contrast $\delta[\bfx(\bfq,t),t] = \det[\partial
x_i/\partial q_j]^{-1} - 1$ is determined by solving the equation of
motion
\begin{eqnarray}
  &&
  \frac{d^2\bfPsi}{dt^2} + 2H\frac{d\bfPsi}{dt} = - \nabla_{x} \Phi(\bfx, 
t),
 \label{1.2} \\
  &&
  \nabla_{x}^{\,2} \Phi = \frac{3}{2} H^2 \Omega \delta(\bfx, t),
\label{1.3}
\end{eqnarray}
where $\nabla_x$ is the spatial derivative with respect to Eulerian
coordinates $\bfx$.  This nonlinear equation for $\bfPsi$ can be
solved by the method of iteration considering
$\partial\mPsi_i/\partial q_j$ to be a small parameter.  The first
iteration corresponds to the ZA.  With further iterations, PZA and
PPZA are considered to improve the accuracy.

There seem to be at least two possible grounds for the Zel'dovich-type
approximations:
\begin{itemize}
\item[(1)] In the plain parallel mass distribution, equations
  (\ref{1.2}) and (\ref{1.3}), with $\delta = 1/(1 + \mPsi') - 1$,
  reduce to
\begin{eqnarray}
  \frac{d^2\mPsi'}{dt^2} + 2H\frac{d\mPsi'}{dt} = \frac32
H^2 \Omega \mPsi',
 \label{1.4}
\end{eqnarray}
where $\mPsi' = \partial \mPsi/\partial q$.  This differential
equation (\ref{1.4}) is exactly the same as that for the linear
density contrast $\delta_{\rm L}$, thus $\mPsi' \propto \delta_{\rm
L}$.Therefore, in this case, ZA becomes exact at least before shell
crossings occur.  This one-dimensional-exact property is considered to
support the validity of this approximation also in three-dimensional
systems.
\item[(2)] ZA is unique in the sense that it is based on the Lagrange
  coordinate scheme while all the other nonlinear approximations are
  based on the Eulerian coordinate scheme.  This fact may make the ZA
  extraordinary excellent.
\end{itemize}
In this paper, we would like to clarify the reason why Zel'dovich-type
approximations are so good from the above limited points of view.

In order to elucidate the aspect (1), we introduce a model of
spheroidal collapse.  In the previous analytical work (Munshi et
al. 1994; Sahni \& Shandarin 1995; Bouchet et al. 1995) only spherical
symmetric models, which are analytically solvable, have been used to
examine the validity of the Zel'dovich-type approximations.  By
changing the axes ratio in our spheroidal model, we can freely control
the effective dimensionality of the system.  For example, a prolate
collapse is effectively dimension two and an oblate collapse is
effectively dimension one.  Therefore, if the one-dimensional-exact
property in the first aspect (1) in the above gives the very reason
for the Zel'dovich-type approximations, their accuracy would be the
best in oblate collapse and may be better in prolate collapse than in
spherical collapse.

In order to explore the aspect (2), we introduce the Pad\'e
approximation in Eulerian coordinate scheme.  This is not a simple
polynomial expansion with a small parameter but a rational polynomial
expansion. In this paper, we compare this sophisticated approximation
in the Eulerian scheme with Zel'dovich-type approximations in
Lagrangian scheme. If the aspect (2) gives the very reasoning for the
Zel'dovich-type approximations, their accuracy is far better than this
Pad\'e approximation.

In the course of our study, we need to establish the hierarchy of
accuracy in various nonlinear approximations.  Therefore we examine
the other known nonlinear approximations in Eulerian scheme such as
(a) linear perturbation and higher order perturbation approximations,
(b) frozen-flow approximation, and (c) linear-potential approximation,
as well as Zel'dovich-type approximations and the Pad\'e
approximation.

The organization of this paper is as follows: In section two, we
summarize various nonlinear approximations in gravitational
instability theory.  They are applied to the spheroidal perturbation
model in section three.  The first aspect is examined in this section.
In section four, we introduce the Pad\'e approximation in Eulerian
scheme.  The second aspect is examined in this section.  In section
five, we clarify which aspects are supported from our analysis.  In
the last section six, we conclude our analysis and briefly mention the
possibility of the approximation scheme beyond the Zel'dovich-type
approximations.

\section{Nonlinear Approximations in Gravitational Instability Theory}
\setcounter{equation}{0}

\subsection{Equations of motion and linear perturbation scheme}

In the gravitational instability theory, the non-relativistic matter
with zero pressure in an Einstein-de Sitter universe is described by
the following set of equations (see e.g., Peebles 1980),
\begin{eqnarray}
   &&
   \dot{\delta} + \nabla\cdot[(1 + \delta)\bfv] = 0,
   \label{eq1}
   \\
   &&
   \dot{\bfv} + 2 H \bfv + (\bfv\cdot\nabla)\bfv +
   \nabla\Phi = \bf0,
   \label{eq2}
   \\
   &&
   \nabla^2 \Phi = \frac32 H^2 \delta,
   \label{eq3}
\end{eqnarray}
where $\bfx$, $\bfv(\bfx,t)$, $\Phi(\bfx,t)$ are respectively
position, peculiar velocity, peculiar potential in comoving
coordinate, which correspond to $a \bfx$, $a \bfv$, $a^2 \Phi$ in
physical coordinate.  An over dot denotes the time derivative and
$\nabla \equiv \partial/\partial\bfx$ denotes the spatial derivative
with respect to comoving coordinates. The scale factor $a$ varies as
$a \propto t^{2/3}$ and the Hubble parameter is $H = \dot{a}/a =
2/(3t)$. Although we consider Einstein-de~Sitter universe throughout
this paper, it is straightforward to generalize our analysis to
$\Omega \neq 1, \Lambda \ne 0$ universes.

For these non-linear equations, we show various approximations in the
following. In the linear regime ($\delta \ll 1$), we can safely
neglect the nonlinear terms and obtain relatively simple solution
(see, e.g., Peebles 1980). Neglecting decaying mode, the solution is
\begin{eqnarray}
\label{eq:a1}
\delta_{\rm L} (\bfx, t) =\frac{a (t)}{a_{\rm in}} \delta_{\rm in} (\bfx), 
\\
\label{eq:a2}
\Phi_{\rm L} (\bfx, t) = \frac{3}{2} H^2 \triangle^{-1} \delta_{\rm L}, \\
\label{eq:a3}
\bfv_{\rm L} (\bfx, t) = - \frac{2}{3 H} \nabla \Phi_{\rm L},
\end{eqnarray}
where $\triangle^{-1}$ is the inverse Laplacian:
\begin{equation}
  \triangle^{-1} F(\bfx) \equiv
  - \frac{1}{4\pi} \int d^3x' \frac{F(\bfx')}{|\bfx' - \bfx|}.
  \label{eq:a4}
\end{equation}
These quantities simply evolve as $\delta_{\rm L} \propto a , \quad
\Phi_{\rm L} \propto a^{-2} ,\quad \bfv_{\rm L} \propto a^{- 1/2}$.

For the later convenience, we introduce the peculiar potential $\phi =
a^2 \Phi$. In Einstein-de~Sitter universe, the linear peculiar
potential is constant, $\phi_{\rm L}(\bfx, t) = {\rm const.} \equiv
\phi (\bfx)$.

\subsection{Higher order perturbation methods in Eulerian scheme}

In higher order perturbation methods, nonlinear correction terms are
added to the linear solution; $\delta = \delta_{\rm L} + \delta^{(2)}
+ \delta^{(3)} + \cdots$, $\Phi = \Phi_{\rm L} + \Phi^{(2)} +
\Phi^{(3)} + \cdots$, $\bfv = \bfv_{\rm L} + \bfv^{(2)} + \bfv^{(3)} +
\cdots$, where $\delta^{(n)}$ $\Phi^{(n)}$ $\bfv^{(n)}$ is assumed to
be of order $(\delta_{\rm L})^n$. This expansion enables us to solve
the nonlinear equations (\ref{eq1})-(\ref{eq3}) order by order. The
second-order expression of the perturbation is relatively simple
(e.g., see Peebles 1980; Fry 1984):
\begin{eqnarray}
  \delta = \delta_{\rm L} + \frac{5}{7} \delta _{\rm
  L}^{\,2} + \delta_{{\rm L},i} \varphi_{{\rm L},i} + \frac{2}{7}
  \varphi_{{\rm L},ij} \varphi_{{\rm L},ij}
  \label{2nddelta}
\end{eqnarray}
where $\varphi = \triangle^{-1}\delta_{\rm L}$.  However, the
expression of higher order is complicated. For the detailed expression
of the third and fourth-order solution in Fourier space, see Goroff et
al. (1986). The explicit expression for the higher order is not
necessary for our analysis, so is not quoted here.

\subsection{Frozen flow and linear potential approximations}

Matarrese et al. (1992) introduced the frozen flow approximation (FF,
hereafter). In this approximation, the velocity field $\bfv(\bfx, t)$
is kept fixed to the value of linear perturbation scheme:
\begin{equation}
\bfv_{\rm FF} \left( \bfx , t \right)
= - \frac{2}{3 a^2 H} \nabla_x \phi_{\rm L} (\bfx), \label{ff}
\end{equation}
A particle in FF moves simply along the line determined by the fixed
linear velocity fields (\ref{ff}), i.e., the position of each
particle, $\bfx(t)$, is described by the differential equation,
\begin{eqnarray}
\frac{d\bfx}{dt}
= - \frac{2}{3 a^2 H} \nabla_x \phi_{\rm L} (\bfx(t)).
\label{eq:a5}
\end{eqnarray}
where $d/dt$ is the Lagrangian time derivative.  It is convenient to
rewrite the equations as follows:
\begin{eqnarray}
\frac{d\bfx}{da}
= - \frac{2}{3 a^3 H^2} \nabla_x \phi_{\rm L} (\bfx).
\label{FF}
\end{eqnarray}

On the other hand, Brainerd, Scherrer \& Villumsen (1993) and Bagla \&
Padmanabham (1994) introduced linear potential approximation (LP,
hereafter) which is based on the assumption that the gravitational
potential evolves according to the linear perturbation scheme. As a
result, particles effectively move along the lines of force of the
initial potential $\phi_{\rm in}$.  Thus the position of each particle
$\bfx(t)$ is described by the differential equation,
\begin{equation}
  \frac{d^2\bfx}{dt^2} +
  2H\frac{d\bfx}{dt} =
  - \frac{1}{a^2} \nabla_x \phi_{\rm L}(\bfx).
  \label{eq:a6}
\end{equation}
This equation is also rewritten as follows:
\begin{equation}
  \frac{d^2\bfx}{da^2} +
  \frac{3}{2a}\frac{d\bfx}{da} +
  \frac{1}{a^4 H^2} \nabla_x \phi_{\rm L}(\bfx) = 0.
  \label{eq:a7}
\end{equation}

\subsection{Zel'dovich approximation and higher order perturbation
methods in Lagrangian scheme}

In the Lagrangian perturbation methods (see, e.g., Bernardeau 1994;
Buchert 1994; Bouchet et al.~1995; Catelan 1995) we consider the
motion of mass elements labelled by unperturbed Lagrangian coordinates
$\bfq$. The comoving Eulerian position of mass element $\bfq$ at time
$t$ is denoted by $\bfx(\bfq,t)$. The displacement field
$\bfPsi(\bfq,t)$ defined by equation (\ref{defpsi}) is the dynamical
variable in this formulation.  Taking divergence and rotation of the
equations of motion (\ref{1.2}), (\ref{1.3}), we obtain the equations
of motion for $\bfPsi(\bfq,t)$,
\begin{eqnarray}
   &&
   \left[\frac{d^2\mPsi_{i,j}}{dt^2} +
     2 H \frac{d\mPsi_{i,j}}{dt} \right]\left(J^{-1}\right)_{ji} +
   \frac32 H^2 \Omega \left(J^{-1} - 1\right) = 0,
   \label{eq2-la4} \\
   &&
   \epsilon_{ijk}
   \left[\frac{d^2\mPsi_{j,l}}{dt^2} +
     2 H \frac{d\mPsi_{j,l}}{dt} \right]\left(J^{-1}\right)_{lk}
   = 0
   \label{eq2-la5}
\end{eqnarray}
where $d/dt$ is Lagrangian time derivative, $J_{ij} = \partial
x_i/\partial q_j = \delta_{ij} + \mPsi_{i,j}$, $J = \det J_{ij}$, and
we used the relation $(\nabla_x)_i = (J^{-1})_{ji}(\nabla_q)_j$.  In
usual treatment of Lagrangian perturbation methods, one assumes an
additional condition, i.e., vorticity-free condition $\nabla_x \times
\bfv = \bf0$, which is equivalent to
\begin{equation}
   \epsilon_{ijk} \frac{d\mPsi_{j,l}}{dt}
   \left(J^{-1}\right)_{lk} = 0,
\label{eq2-la6}
\end{equation}
which replaces equation (\ref{eq2-la5}).  The solutions with this
vorticity-free condition form a subclass of the all general solutions
[rotational perturbation is argued by Buchert (1992) and Buchert \&
Ehlers (1993)].  Therefore, equations (\ref{eq2-la4}) and
(\ref{eq2-la6}) are solved perturbatively for derivatives of
displacement field $\mPsi_{i,j} = \mPsi_{i,j}^{(1)} +
\mPsi_{i,j}^{(2)} + \mPsi_{i,j}^{(3)} + \cdots$, keeping only terms of
leading time-dependence [see Buchert \& Ehlers (1993), Buchert (1994),
Bouchet et al.~(1995), and Munshi et al.~(1994) for detail]. In
Einstein-de~Sitter universe, the time dependence of each terms is
separated from its spatial dependence:
\begin{equation}
   \bfPsi^{(n)} = \left(\frac{2}{3 a^2 H^2}\right)^n \bfpsi^{(n)}
    \left( \bfq \right)
   \label{eq2-la7}
\end{equation}
The resulting perturbative solutions, up to third-order, are
\begin{eqnarray}
&&
\psi^{(1)}_i = - \partial_i \phi_{\rm L} (\bfq),
\label{eq:m01}
\\
&&
\psi^{(2)}_i = - \frac{3}{14} \partial_i \triangle^{-1}
\left(  \psi^{(1)}_{j,j} \psi^{(1)}_{k,k}
  - \psi^{(1)}_{j,k} \psi^{(1)}_{j,k} \right),
\label{eq:m02}
\\
&&\psi^{(3)}_i = - \frac{5}{9} \partial_i \triangle^{-1}
\left( \psi^{(1)}_{j,j} \psi^{(2)}_{k,k}
   - \psi^{(1)}_{j,k} \psi^{(2)}_{k,j} \right)
- \frac{1}{3} \partial_i \triangle^{-1}
{\rm det} \left[ \psi^{(1)}_{j,k} \right]
\nonumber \\
&&\qquad\qquad
 - \frac{1}{3} \partial_j \triangle^{-1}
   \left( \psi^{(1)}_{k,j} \psi^{(2)}_{i,k}
   - \psi^{(1)}_{k,i} \psi^{(2)}_{j,k} \right).
\label{eq:m03}
\end{eqnarray}
The first-order solution is equivalent to the ZA.  While the first-
and second-order displacement field is irrotational, the third-order
solution has both the rotational part (the last term) and the
irrotational part (first two terms).

\section{Spheroidal perturbation}
\setcounter{equation}{0}

In this section, we examine our aspect (1), one-dimensional exact
property of Zel'dovich-type approximations applying various non-linear
approximations summarized in the previous section to the spheroidal
collapse model.

\subsection{Equation of motion of an ellipsoid}

A particle motion in an Einstein-de~Sitter universe is described by
the following equation of motion:
\begin{eqnarray}
  &&
  \frac{d^2\bfx}{dt^2} + 2H\frac{d\bfx}{dt} = - \nabla_{x} \Phi(\bfx, t),
  \label{eq:m1} \\
  &&
  \nabla_{x}^{\,2} \Phi = \frac{3}{2} H^2 \delta(\bfx, t),
  \label{eq:m2}
\end{eqnarray}
where $d/dt$ is the Lagrangian time derivative as before. In a
homogeneous ellipsoid, the density perturbation $\delta(\bfx,t)$ is
given by
\begin{equation}
  \delta(\bfx, t) =
  \delta_{\rm e}(t)\,\,\,
  \Theta\left(
    1 - 
    \frac{x_1^{\,2}}{\alpha_1^{\,2}(t)} -
    \frac{x_2^{\,2}}{\alpha_2^{\,2}(t)} -
    \frac{x_3^{\,2}}{\alpha_3^{\,2}(t)}
  \right),
  \label{eq:m3}
\end{equation}
where $\alpha_i$ are the half-length of the principal axes of the
ellipsoid and $\Theta$ is a step function.  The solution of the
Poisson equation (\ref{eq:m2}) inside the homogeneous ellipsoid is
known (see, e.g., Kellogg 1953; Binney \& Tremaine 1987),
\begin{eqnarray}
&&
  \triangle^{-1}
  \Theta\left(
    1 - 
    \frac{x_1^{\,2}}{\alpha_1^{\,2}} -
    \frac{x_2^{\,2}}{\alpha_1^{\,2}} -
    \frac{x_3^{\,2}}{\alpha_1^{\,2}}
  \right) =
  \frac{1}{4}
  \sum^{3}_{i=1} A_i x_i^{\,2},
  \label{eq:m3.5}
\\
&&
  \qquad\qquad\qquad 
  ({\it r.h.s.}\  {\rm is\ only\ for\ inside\ the\ ellipsoid})
\nonumber
\end{eqnarray}
where
\begin{equation}
  A_i = \alpha_1\alpha_2\alpha_3
  \int_0^\infty (\alpha_i^{\,2} + \lambda)^{-1}
  \prod_{j=1}^3
  (\alpha_j^{\,2} + \lambda)^{-1/2} d\lambda.
  \label{eq:m5}
\end{equation}
These coefficients $A_i$ automatically satisfy the following constraint 
\begin{equation}
  \sum_{i=1}^3 A_i = 2.
  \label{eq:m6}
\end{equation}
Thus, the solution of equation (\ref{eq:m2}) becomes
\begin{equation}
  \Phi = \frac{3}{8} H^2 \delta_{\rm e}
  \sum_{i=1}^3 A_i(t) x_i^{\,2}.
  \label{eq:m4}
\end{equation}

The quadratic form of the potential (\ref{eq:m4}) implies that there
exists a solution that retain homogeneity of the ellipsoid if the
initial velocity field is linear in space and if the homogeneity of
outer region is assumed. (Lynden-Bell 1962; 1964; Lin, Mestel \& Shu
1965; Icke 1973)

Combining the equations (\ref{eq:m1}) and (\ref{eq:m4}),we obtain
\begin{equation}
  \ddot{x}_i + 2H \dot{x}_i =
  - \frac{3}{4} H^2 \delta_{\rm e} A_i x_i,
  \label{eq:m7}
\end{equation}
which describes the motion of the particles inside the ellipsoid.
Generally, the variables $\delta_{\rm e}$ and $A_i$'s depend on the
position of other particles, i.e., volume and shape of the ellipsoid.
Fortunately in our model, it is sufficient to consider only three
particles at the coordinates $(\alpha_1(t),0,0)$, $(0, \alpha_2(t),0)$
and $(0,0,\alpha_3(t))$ because these three particles completely
characterize the motion of the entire ellipsoid. The density contrast
of the ellipsoid is given by
\begin{equation}
  \delta_{\rm e} =
  \frac{a^3}{a_{\rm in}^{\,3}}
  (1 + \delta_{\rm in})
  \frac{\alpha_{\rm in1}\alpha_{\rm in2}\alpha_{\rm in3}}
    {\alpha_1 \alpha_2 \alpha_3}
  - 1,
  \label{eq:m8}
\end{equation}
where $\alpha_{\rm in} = \alpha(t_{\rm in})$, $a_{\rm in} = a(t_{\rm
in})$ and $t_{\rm in}$ is the initial time. The equation of motion for
the three points are given by
\begin{equation}
  \ddot{\alpha}_i + 2H \dot{\alpha}_i =
  - \frac{3}{4} H^2 \delta_{\rm e} A_i \alpha_i.
  \label{eq:m9}
\end{equation}
The above equations (\ref{eq:m5}), (\ref{eq:m8}) and (\ref{eq:m9}) are
the closed set of equations of motion which describe the motion of the
entire ellipsoid. Because of the simplicity, the homogeneous ellipsoid
model can be used as an approximation for protoobjects in structure
formation in the universe (White \& Silk 1979; Eisenstein \& Loeb
1995; Bond \& Myers 1996).

We numerically solve these equations. For the initial condition of the
numerical integration, we adopt $\delta_{\rm in} = 10^{-5}$--$10^{-9}$
and the ZA for velocity field. For sufficiently small value of
$\delta_{\rm in}$, it is accurate enough to adopt ZA to prepare the
initial velocity field.  Actually, adopting PZA or PPZA does not
change the results and the simple ZA is sufficient.  In this paper, we
adopt spheroidal symmetry, $\alpha_1 = \alpha_2$, in which case,
$A_i$'s have analytic forms as follows (Peebles 1980):
\begin{equation}
   A_1 = A_2 = \frac{2}{3} (1 + h), \qquad
   A_3 = \frac{2}{3} (1 - 2 h),
   \label{eq:m10}
\end{equation}
where
\begin{equation}
  h =
  \left\{
    \begin{array}{lll}
      \displaystyle
      \frac{3}{2} \frac{\sqrt{1 - e^2}}{e^3} \sin^{-1} e -
      \frac{3 - e^2}{2 e^2},
      &
      \displaystyle
      e = \sqrt{1 - \left(\frac{\alpha_3}{\alpha_1}\right)^2}
      &
      \displaystyle
      \left(\alpha_1 = \alpha_2 > \alpha_3\right)
      \\[0.3cm]
      \displaystyle
      \frac34 \frac{1 - \bar{e}^2}{\bar{e}^3}
      \ln\left(\frac{1 - \bar{e}}{1 + \bar{e}}\right) +
      \frac{3 - \bar{e}^2}{2 \bar{e}^2},
      &
      \displaystyle
      \bar{e} = \sqrt{1 - \left(\frac{\alpha_1}{\alpha_3}\right)^2}
      &
      \displaystyle
      \left(\alpha_1 = \alpha_2 < \alpha_3\right)
   \end{array}
   \right.
   \label{eq:m11}
\end{equation}
This spheroidal symmetric model is sufficient to study the dimensionality
of the system.
\subsection{Linear perturbation scheme}

The evolution of spheroidal perturbations in linear perturbation
scheme is simple:
\begin{equation}
  \delta_{\rm L}(\bfx,t) =
      \frac{a}{a_{\rm in}} \delta_{\rm in}
      \,\,\,\Theta\left(
        1 -
        \frac{x_1^{\,2}}{\alpha_{{\rm in}1}^{\,2}} -
        \frac{x_2^{\,2}}{\alpha_{{\rm in}2}^{\,2}} -
        \frac{x_3^{\,2}}{\alpha_{{\rm in}3}^{\,2}}
      \right),
\end{equation}
which reduces to
\begin{equation}
  \delta_{\rm L} (\bfx,t) =
      \pm a
      \,\,\,\Theta\left(
        1 -
        \frac{x_1^{\,2}}{\alpha_{{\rm in}1}^{\,2}} -
        \frac{x_2^{\,2}}{\alpha_{{\rm in}2}^{\,2}} -
        \frac{x_3^{\,2}}{\alpha_{{\rm in}3}^{\,2}}
      \right),
 \label{L}
\end{equation}
after the normalization of scale factor $a$ as $a_{\rm in} =
|\delta_{\rm in}|$.  Here and after, upper sign corresponds to the
positive perturbations $\delta_{\rm e} > 0$, and the lower sign
corresponds to the negative perturbations $\delta_{\rm e} <
0$. Solving Poisson equation by equation (\ref{eq:m3.5}), linear
peculiar potential inside the spheroid is given by
\begin{eqnarray}
\phi_{\rm L} (\bfx) = \pm \frac{a^3 H^2}{4}
\left[ (1 + h_{\rm in}) (x_1^2 + x_2^2)
+ (1 - 2 h_{\rm in}) x_3^2 \right].
\label{linearpot}
\end{eqnarray}
In Figures 1--5, we plot $\delta_{\rm L}$ against the numerically
solved true density contrast $\delta_{\rm true}$ for various
axis-ratios and for positive and negative perturbations.

\subsection{Frozen flow approximation}

We now derive the evolution of spheroidal perturbation in FF. Linear
peculiar potential inside the spheroid is given by equation
(\ref{linearpot}) and the differential equation (\ref{FF}) reduces to
\begin{equation}
\frac{\partial x_{(1,2)}}{\partial a}
= \mp \frac{1}{3} (1 + h_{\rm in}) x_{(1,2)}, \quad
\frac{\partial x_3}{\partial a}
= \mp \frac{1}{3} (1 - 2 h_{\rm in}) x_3.
\end{equation}
where (1,2) denotes the subscript 1 or 2 in this order. The solution
of these equations are
\begin{equation}
x_{(1,2)} = q_{(1,2)} \exp \left( \mp \frac{1 + h_{\rm in}}{3} a \right), 
\quad
x_3 = q_3 \exp \left( \mp \frac{1 - 2 h_{\rm in}}{3} a \right),
\end{equation}
where $q_i$'s are constants of integration which correspond to the
initial condition, $x_i \rightarrow q_i$ as $a \rightarrow 0$. Because
$\delta \rightarrow 0$ as $a \rightarrow 0$, and $\rho \propto (x_1
x_2 x_3)^{-1}$, evolution of the density contrast is
\begin{eqnarray}
\delta = \frac{q_1 q_2 q_3}{x_1 x_2 x_3} - 1
= \exp (\pm a) - 1. \label{LPdelta}
\end{eqnarray}
It should be noted that the solution (\ref{LPdelta}) does not depend
on the parameter $h_{\rm in}$, and has the same evolution as in the
spherical model, $h_{\rm in} = 0$.  In Figures 1--5, we plot the
evolution of FF with thin dot-dotted dash lines against $\delta_{\rm
true}$.

\subsection{Linear potential approximation}

Similarly, the solution of LP for homogeneous spheroids is obtained as
follows. Equations (\ref{eq:a7}) and (\ref{linearpot}) imply the
equation of motion of particles in LP:
\begin{eqnarray}
\left\{ \begin{array}{lll}
\displaystyle\frac{d^2 x_{(1,2)}}{d a^2} +
\displaystyle\frac{3}{2 a} \frac{d x_{(1,2)}}{d a}
\pm \displaystyle\frac{1}{2 a} (1 + h_{\rm in}) x_{(1,2)}
 = 0, \\[0.2cm]
\displaystyle\frac{d^2 x_3}{d a^2} + \frac{3}{2 a} \frac{d x_3}{d a}
\pm \frac{1}{2 a} (1 - 2 h_{\rm in}) x_3
 = 0.
\end{array}   \right.
\end{eqnarray}
The solutions of these equations are
\begin{eqnarray}
& &_{} \left\{ \begin{array}{lll}
x_{(1,2)} &=&
  \displaystyle\frac{q_{(1,2)}}{\sqrt{2 (1 + h_{\rm in}) a}}
  \sin \sqrt{2 (1 + h_{\rm in}) a} \\[0.2cm]
x_3 &=&
  \displaystyle\frac{q_3}{\sqrt{2 (1 - 2 h_{\rm in}) a}}
  \sin \sqrt{2 (1 - 2 h_{\rm in}) a}
\end{array}   \right. \; \; (\delta > 0), \\
& &_{} \left\{ \begin{array}{lll}
x_{(1,2)} &=& \displaystyle\frac{q_{(1,2)}}{\sqrt{2 (1 + h_{\rm in}) a}}
  \sinh \sqrt{2 (1 + h_{\rm in}) a} \\[0.2cm]
x_3 &=& \displaystyle\frac{q_3}{\sqrt{2 (1 - 2 h_{\rm in}) a}}
  \sinh \sqrt{2 (1 - 2 h_{\rm in}) a}
\end{array}   \right. (\delta < 0).
\end{eqnarray}
The evolution of density contrast is, therefore,
\begin{eqnarray}
\delta = \left\{ \begin{array}{rr}
\displaystyle\frac{2 \sqrt{2} (1 + h_{\rm in})
(1 - 2 h_{\rm in})^{1/2} a^{3/2}}
{\sin^2 \sqrt{2 (1 + h_{\rm in}) a} \sin \sqrt{2 (1 - 2 h_{\rm in}) a}}
- 1 & (\delta > 0) \\[0.5cm]
\displaystyle\frac{2 \sqrt{2} (1 + h_{\rm in})
  (1 - 2 h_{\rm in})^{1/2} a^{3/2}}
{\sinh^2 \sqrt{2 (1 + h_{\rm in}) a} \sinh \sqrt{2 (1 - 2 h_{\rm in}) a}}
- 1 & (\delta < 0)
\end{array}   \right. .
\end{eqnarray}
In the spherical model ($h_{\rm in}$ = 0), the expression reduces to
\begin{eqnarray}
\delta = \left\{ \begin{array}{rr}
\left(\displaystyle\frac{\sqrt{2 a}}{\sin \sqrt{2 a}} \right)^3
- 1 & (\delta > 0)\\[0.5cm]
\left(\displaystyle\frac{\sqrt{2 a}}{\sinh \sqrt{2 a}} \right)^3
- 1 & (\delta < 0)
\end{array}   \right. ,
\end{eqnarray}
which corresponds to the result already derived by Brainerd et
al. (1993). In Figures 1--5, we plot the evolution of LP with thin
short-dashed lines against $\delta_{\rm true}$.

\subsection{Zel'dovich approximation and higher order Lagrangian
perturbation methods}\label{ZTAs}

In ZA, the particle motion is described by the equation
(\ref{eq:m01}). Using equation (\ref{linearpot}) for spheroidal case,
we obtain
\begin{eqnarray}
& &{} \left\{ \begin{array}{lll}
\psi^{(1)}_{(1,2)} = \mp \displaystyle\frac{a^3 H^2 q_{(1,2)}}{2}
  (1 + h_{\rm in}) \\[0.2cm]
\psi^{(1)}_3 = \mp \displaystyle\frac{a^3 H^2 q_3}{2} (1 - 2 h_{\rm in})
\end{array}   \right. .
\label{1stelidis}
\end{eqnarray}

As for the second-order term (\ref{eq:m02}), the inverse Laplacian can
be calculated by equation (\ref{eq:m3.5}), because we assume that
there is no fluctuation outside the spheroid all the time, i.e.,
$\delta = 0$ outside the spheroid in q-space.  Thus in the Lagrangian
perturbation scheme, we can easily identify the boundary of the
spheroid.  This is the point that the Lagrangian perturbation scheme
is technically advantageous than the Eulerian perturbation scheme.  We
will return to this argument later.  Thus equation (\ref{eq:m02})
implies
\begin{eqnarray}
\left\{ \begin{array}{lll}
\psi^{(2)}_{(1,2)}
= - \displaystyle\frac{3 a^6 H^4 q_{(1,2)}}{28} (1 + h_{\rm in}
- h_{\rm in}^2 - h_{\rm in}^3), \\[0.2cm]
\psi^{(2)}_3
= - \displaystyle\frac{3 a^6 H^4 q_3}{28} (1 - 2 h_{\rm in}
- h_{\rm in}^2 + 2 h_{\rm in}^3).
\end{array} \right.
\label{2ndelidis}
\end{eqnarray}

Similarly, we can calculate the third-order terms according to the
equation (\ref{eq:m03}):
\begin{eqnarray}
& &_{} \left\{ \begin{array}{lll}
\psi^{(3)}_{(1,2)} =
\mp \displaystyle\frac{a^9 H^6 q_{(1,2)}}{504} \left( 23 + 23  h_{\rm in}
- 39  h_{\rm in}^2 - 25  h_{\rm in}^3 + 44 h_{\rm in}^4
+ 30  h_{\rm in}^5 \right) \\[0.2cm]
\psi^{(3)}_3 =
\mp \displaystyle\frac{a^9 H^6 q_3}{504} \left( 23 - 46  h_{\rm in}
- 39  h_{\rm in}^2 + 92  h_{\rm in}^3 + 2 h_{\rm in}^4
- 60  h_{\rm in}^5  \right)
\end{array}   \right. .
\label{3rdelidis}
\end{eqnarray}

The density contrast in these Zel'dovich-type approximations is given
by
\begin{equation}
\delta = \frac{1}{(1 + \mPsi_1/q_1)(1 + \mPsi_2/q_2)(1 + \mPsi_3/q_3)}
 - 1, \label{yuragi}
\end{equation}
where $\mPsi_{i} = \mPsi^{(1)}_{i}$ for ZA, $\mPsi_{i} =
\mPsi^{(1)}_{i}+\mPsi^{(2)}_{i}$ for PZA, and $\mPsi_{,i} =
\mPsi^{(1)}_{i}+\mPsi^{(2)}_{i}+\mPsi^{(3)}_{i}$ for PPZA. The
relation between $\psi$ and $\mPsi$ is given by (\ref{eq2-la7})

In the spherical case, $h_{\rm in} = 0$, the equation (\ref{yuragi})
reduces to
\begin{eqnarray}
&&
\delta =
\left( 1 \mp \displaystyle\frac{a}{3} \right)^{-3} - 1,
\\
&&
\delta =
\left( 1 \mp \displaystyle\frac{a}{3} - \frac{a^2}{21} \right)^{-3} - 1,
\\
&&
\delta =
\left( 1 \mp \displaystyle\frac{a}{3} - \frac{a^2}{21}
    \mp \frac{23 a^3}{1701} \right)^{-3} - 1,
\label{eq:mm1}
\end{eqnarray}
for, respectively, ZA, PZA and PPZA, which correspond to the result by Munshi
et al. (1994). The Lagrangian perturbation methods for ellipsoidal
collapse with respect to mass function is considered by Monaco (1997).
\par
In Figures 1-5, we plot the evolution of ZA,
PZA and PPZA against $\delta_{\rm true}$.

\subsection{Second- and third-order perturbation methods in Eulerian
scheme}

In contrast to Lagrangian perturbation methods, the surface of the
spheroid cannot be explicitly expressed in Eulerian perturbation
methods. This problem makes the calculation of the spheroidal
perturbation difficult in Eulerian perturbation methods. We circumvent
this difficulty by transforming the expression already obtained in
Lagrangian perturbation scheme to that in Eulerian perturbation
scheme.  In doing so, we compare the small expansion parameter in both
schemes.  It is $\Psi_{i,j}$ in Lagrangian perturbation scheme and the
density contrast $\delta$ in Eulerian perturbation scheme.  We notice
these two parameters are the same order, i.e., $\delta \sim {\cal
O}(\Psi_{i,j})$.  We also notice the $n$-th order perturbative
solution in Eulerian scheme, $\delta^{(n)}$ is of order $(\delta_{\rm
L})^n$ and $\Psi^{(n)}_{i,j} \sim {\cal O}(\Psi^{(1)}_{i,j})^n$ in
Lagrangian scheme.  Therefore $\Psi^{(n)}_{i,j} \ \sim \delta^{(n)}$.

Thus, to obtain the perturbative series up to $n$-th order of density
contrast in Eulerian scheme, it is sufficient to expand the density
contrast of $n$-th order in Lagrangian scheme by parameters
$\Psi^{(1)}_{i,j}$ and re-express it in terms of $\delta_{\rm L}$. In
our case, $\Psi^{(n)} \propto a^n$, so Eulerian perturbative series
can be simply obtained by expanding equation (\ref{yuragi}) in terms
of expansion factor $a$. The result is
\begin{eqnarray}
\delta &=& \pm a + \left( \frac{17}{21} +
 \frac{4}{21} h_{\rm in}^2\right) a^2
\pm \left(\frac{341}{567} + \frac{74}{189} h_{\rm in}^2 -
  \frac{4}{81} h_{\rm in}^3 - \frac{8}{189} h_{\rm in}^4\right) a^3.
\label{eulerexp}
\end{eqnarray}

Although we adopt the above method in this paper, this is not the
unique choice.  Actually, we can straightforwardly obtain the Eulerian
perturbation expansion independently from the Lagrangian perturbation
scheme. However in the Eulerian scheme, our assumption `` there is no
fluctuation outside of the spheroid'' becomes ambiguous.  This is
because, in Eulerian coordinate scheme, the location of the surface of
the spheroid cannot be explicitly described as mentioned above.  In
Lagrangian scheme, we assumed that there is no fluctuation outside the
{\em evolved} surface of the spheroid in section \ref{ZTAs}.  In
Eulerian space, on the other hand, it seems technically favorable to
assume that there is no fluctuation outside the {\em initial} surface
of the spheroid.  This is because the inverse Laplacian
(\ref{eq:m3.5}) is easily solved for the latter assumption but is
difficult for the former assumption.  The sole ambiguity in our model
comes from how to fix the fluctuation outside the spheroid.  This
fixing is necessary for us to obtain the numerically exact solutions
for the evolution of density contrast.
For comparison, we quote the result adopting latter assumption
in Eulerian space, no fluctuation outside
the {\em initial} suffice of the spheroid:
\begin{eqnarray}
\delta &=& \pm a + \left( \frac{17}{21} +
  \frac{4}{21} h_{\rm in}^2 \right) a^2
\pm \left(\frac{341}{567} + \frac{38}{105} h_{\rm in}^2 -
 \frac{8}{405} h_{\rm in}^3 + \frac{16}{945} h_{\rm in}^4 \right) a^3.
\label{eulerexp2}
\end{eqnarray}

These two expressions above are very similar in the following two
aspects. Firstly the spherical terms (i.e., terms independent of
$h_{\rm in}$) are the same in both expressions.  This is because for
the spherical perturbation, the evolution of density contrast at a
point is determined only by the inside of the spherical shell on which
the point is located; the artificial fixing of fluctuations outside
the shell does not affect the evolution of density contrast at the
point.  Secondly the two expressions agree with each other up to the
second-order. This can be understood as follows. In linear
perturbation scheme, the artificial fixing of fluctuation outside the
spheroid is not necessary simply because $\delta \propto a$ and the
perturbation is uniform in the spheroid. In second-order methods, the
fluctuation outside the spheroid does not generally vanish however we
force it to vanish artificially. This artificial fixing does not
affect the evaluation of the density contrast inside of spheroid
within the second-order perturbation scheme because the second-order
density contrast is determined as the non-local functional of density
contrast of linear perturbation scheme. The artificial fixing,
however, affects the evaluation of the third and higher order
calculations. Thus the artificial fixing of fluctuations results in
the difference of third and higher order perturbation terms.

In Figures 1-5, we plot the evolution of second- and third-order
result (\ref{eulerexp} thin graph) against $\delta_{\rm true}$.

\begin{figure}
   \leavevmode\psfig{figure=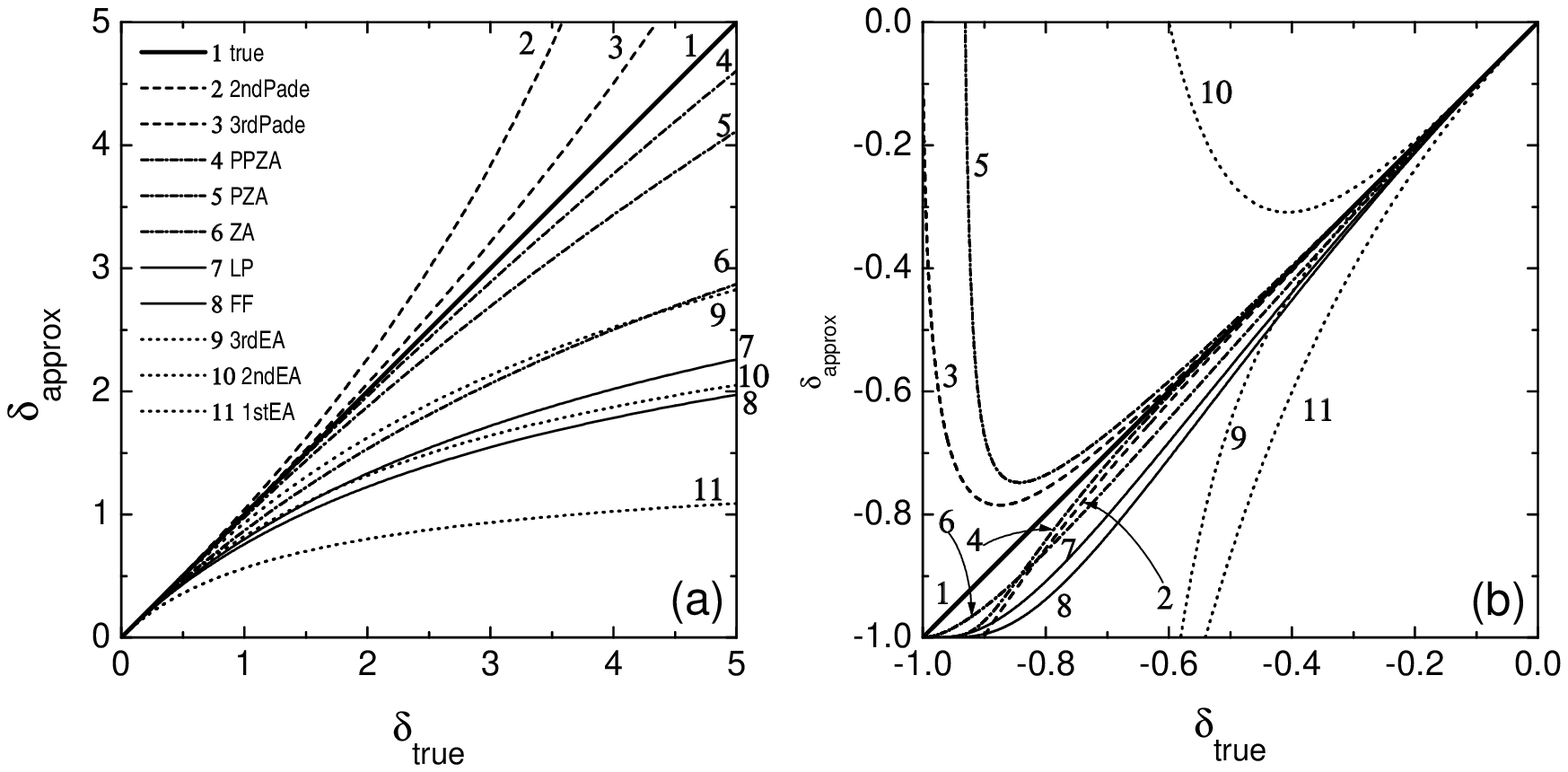,width=15.8cm}
\caption{Density contrast in various approximations for
spherical density perturbations.  Vertical axis is the density
contrasts $\delta_{\rm approx}$ of various nonlinear approximations
and the horizontal axis is the true solution $\delta_{\rm true}$ in
spherical collapse model.  The true solution corresponds to the
diagonal straight line.  The distance to this line represents the
accuracy of each approximation.  Fig. 1(a) is for a case of positive
perturbation and Fig. 1(b) is for a case of negative
perturbation. Some of lines in this figure is previously appeared in
Munshi et al. (1994) and Sahni \& Shandarin (1996). \label{fig1}}
\end{figure}
\begin{figure}
\begin{center}
   \leavevmode\psfig{figure=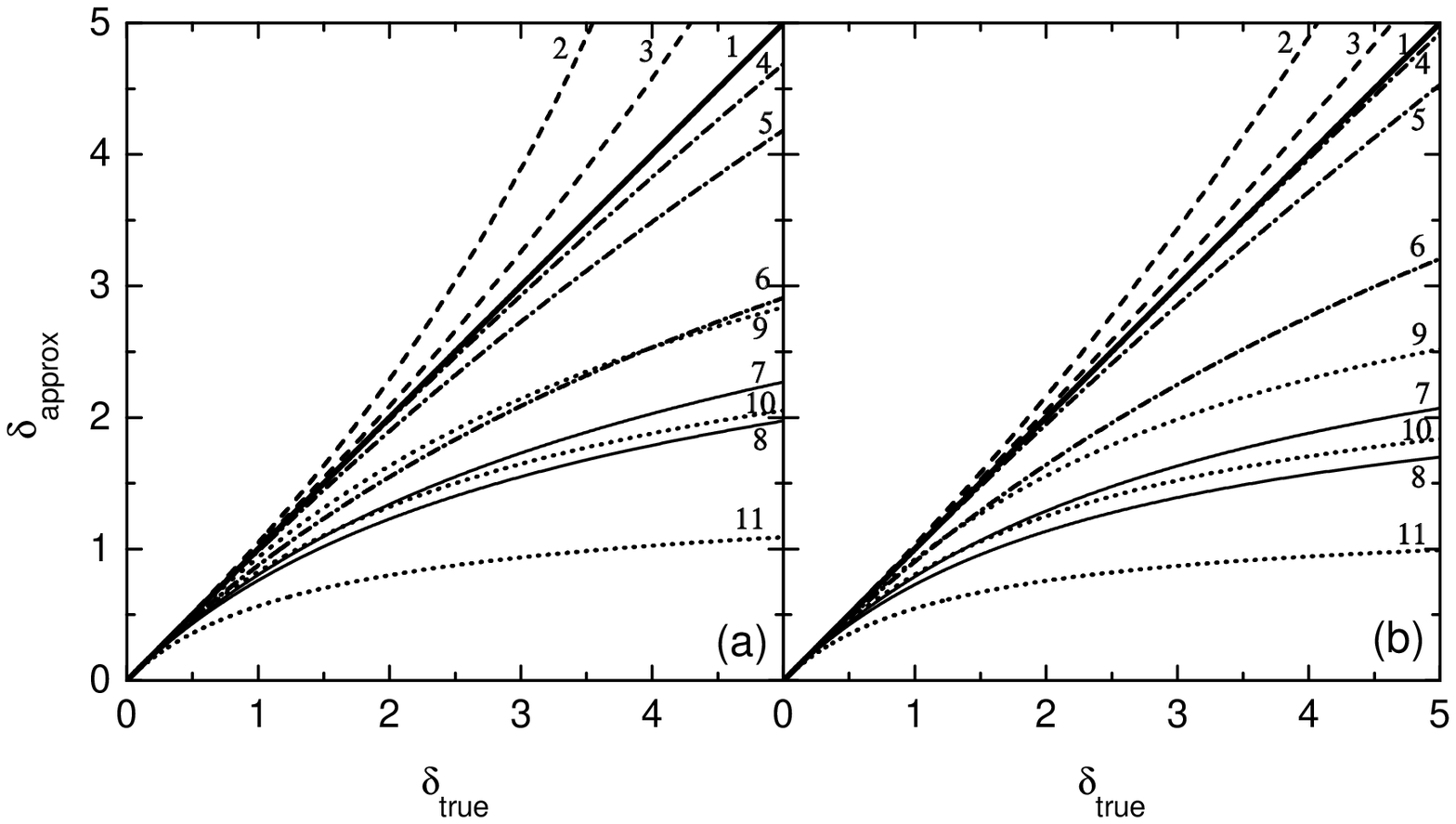,width=15.8cm}
\end{center}
\caption{Density contrast in various approximations for
oblate density perturbations.  The same figure as in Fig. 1, but for
oblate spheroidal collapse. Fig. 2(a) is the case with the initial
axis-ratio $\alpha_{\rm in3}/\alpha_{\rm in1} = 0.8$, and Fig. 2(b) is
with $\alpha_{\rm in3}/\alpha_{\rm in1} = 0.3$. \label{fig2}}
\end{figure}
\begin{figure}
\begin{center}
   \leavevmode\psfig{figure=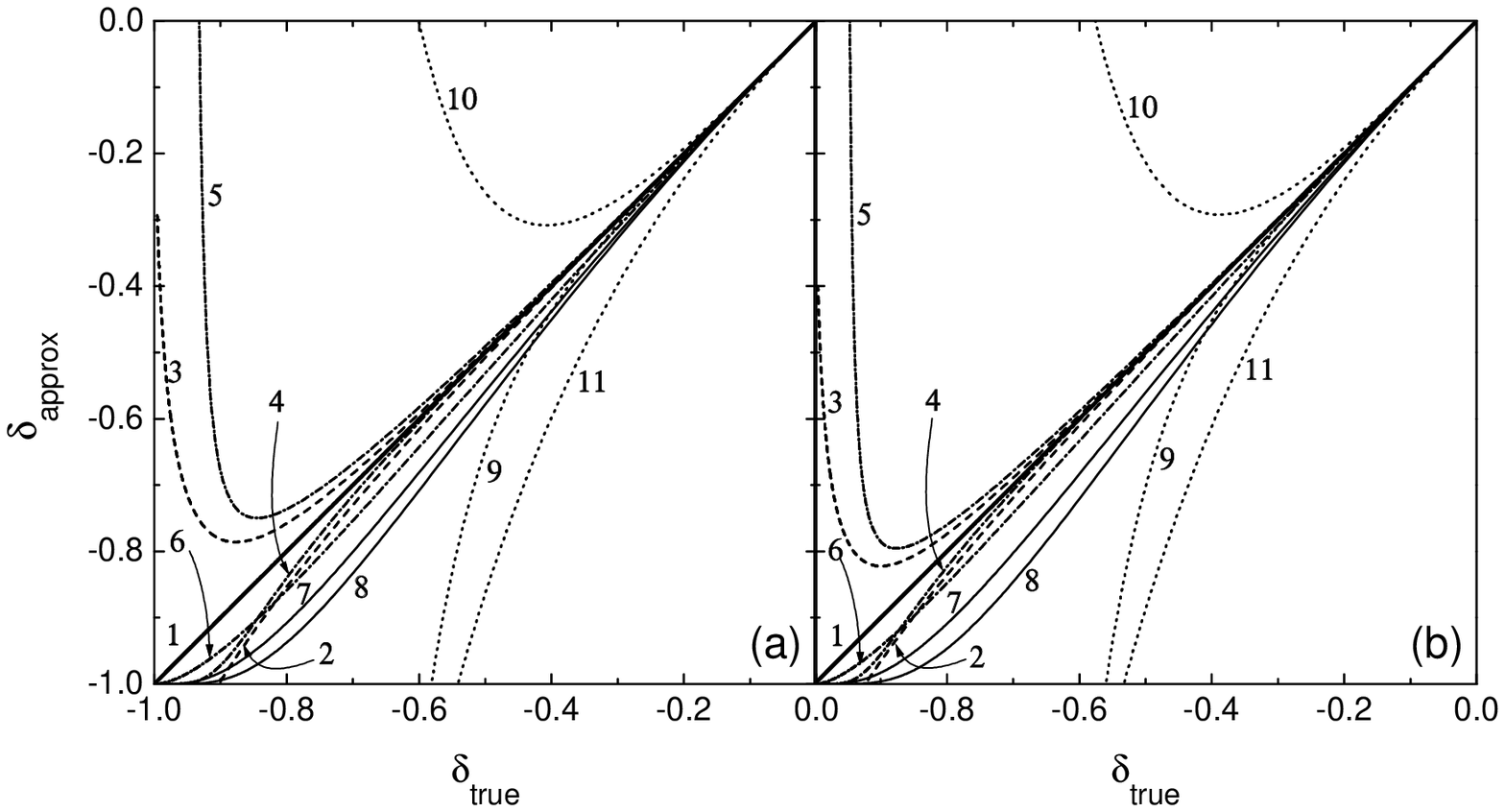,width=15.8cm}
\end{center}
\caption{Density contrast in oblate spheroidal collapse of
negative perturbation.  Same figure as Fig. 2 but for negative
perturbations.  Fig. 3(a) is the case with the initial axis-ratio
$\alpha_{\rm in3}/\alpha_{\rm in1} = 0.8$, and Fig.3(b) is with
$\alpha_{\rm in3}/\alpha_{\rm in1} = 0.3$. \label{fig3}}
\end{figure}
\begin{figure}
\begin{center}
   \leavevmode\psfig{figure=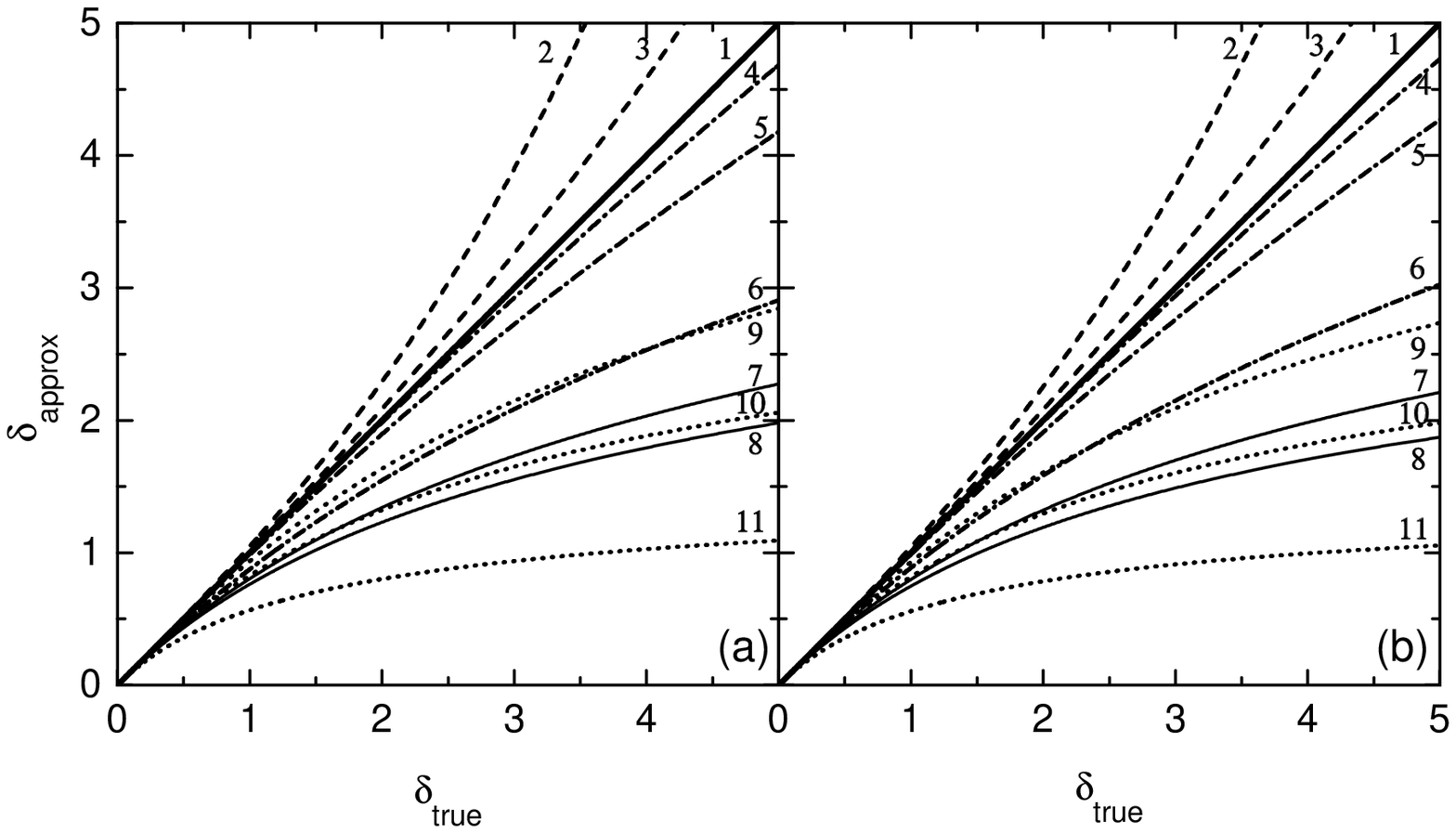,width=15.8cm}
\end{center}
\caption{Density contrast in prolate spheroidal collapse
of positive perturbation.  The same figure as in Fig. 2, but for
prolate spheroidal collapse. Fig. 4(a) is the case with the initial
axis-ratio $\alpha_{\rm in3}/\alpha_{\rm in1} = 1.2$, and Fig. 4(b) is
with $\alpha_{\rm in3}/\alpha_{\rm in1} = 3$. \label{fig4}}
\end{figure}
\begin{figure}
\begin{center}
   \leavevmode\psfig{figure=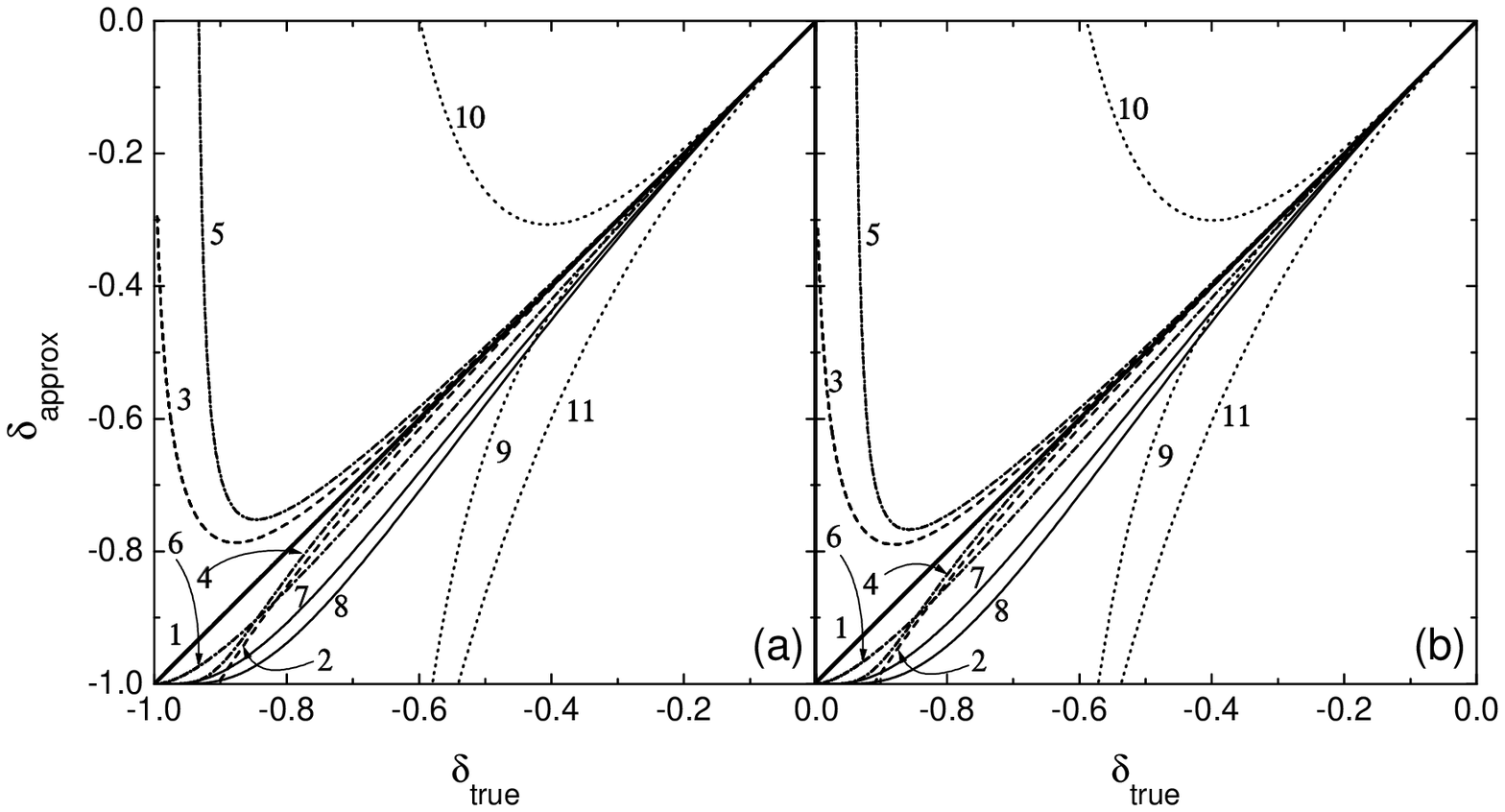,width=15.8cm}
\end{center}
\caption{Density contrast in prolate spheroidal collapse
of negative perturbation.  Same figure as Fig. 3 but for prolate
spheroidal collapse.  Fig. 5(a) is the case with the initial
axis-ratio $\alpha_{\rm in3}/\alpha_{\rm in1} = 1.2$, and Fig.5(b) is
with $\alpha_{\rm in3}/\alpha_{\rm in1} = 3$. \label{fig5}}
\end{figure}

\section{Pad\'{e} approximation}
\setcounter{equation}{0}

In this section, we examine the aspect (2), Lagrange scheme, for the
validity of Zel'dovich-type approximations.  The Zel'dovich-type
approximations are unique in the sense that they are grounded on the
Lagrangian coordinate scheme.  It is advantageous to use this
Lagrangian scheme because the inertia term is linearized in velocity.
Is Lagrangian scheme the indispensable reason why Zel'dovich-type
approximations work accurately beyond the linear regime?  For the
purpose of addressing this issue, we introduce now the Pad\'e
approximation (see, e.g., Press et al. 1992) in Eulerian coordinate
scheme.  This is an approximation for some unknown underlying function
$f(x)$ with rational function whose power series expansion agrees with
a given power series of $f(x)$ to the highest possible order.  Pad\'e
approximation can even simulate the poles of the underlying function
$f(x)$ and is generally better than simple polynomial approximations.

Pad\'{e} approximation for a given unknown function $f(x)$ is
expressed as the ratio of two polynomials.
\begin{eqnarray}
R (x) \equiv \frac{\displaystyle \sum_{k = 0}^{M} a_k x^k}
{1 + \displaystyle \sum_{k = 1}^{N} b_k x^k},
\end{eqnarray}
where $a_k$ and $b_k$ are constant coefficients.  Suppose we already
know the first $M+N$ coefficients $c_k$ of a series expansion of the
function $f(x)$ around $x=0$.
\begin{eqnarray}
\sum_{k = 0}^{M + N} c_k x^k.
\end{eqnarray}
Then the above coefficients $a_k$ and $b_k$ are determined so that the
first $M+N$ coefficients of a series expansion of $R(x)$ agree with
the coefficients $c_k$.\footnote{ Choices $N=M$ or $N=M+1$ are usually
adopted.
We used $N=M+1$ in this paper.  %
}
This condition
\begin{eqnarray}
c_0 = a_0, \nonumber \\
\sum_{m = 1}^{N} b_m c_{N - m + k} &=&
   - c_{N + k},\qquad k = 1,\cdots,N \nonumber \\
\sum_{m = 0}^{k} b_m c_{k - m} &=&
   a_k,\qquad \qquad k = 1,\cdots,N \nonumber
\label{padecoeff}
\end{eqnarray}
fixes the coefficients $a_k$ and $b_k$.

Now we use Pad\'{e} approximation with the perturbative expansions in
Eulerian coordinate scheme.  The density contrast in the spheroidal
model up to the third-order perturbation is given by equation
(\ref{eulerexp}). The corresponding Pad\'{e} approximation is given by
\begin{eqnarray}
\delta &=& \frac{\pm a}{ 1
\mp \displaystyle\frac{17 + 4 h_{\rm in}^2}{21} a
+ \left(\displaystyle\frac{214}{3969} - \frac{110}{1323} h_{\rm in}^2
+ \frac{4}{81} h_{\rm in}^3 + \frac{104}{1323} h_{\rm in}^4 \right)
a^2} .
\label{eq:mm2}
\end{eqnarray}
The second-order Pad\'e approximation is obtained discarding the $a^2$
term in the denominator of the above equation.

These approximations are already superposed on top of the previous
approximations in Figures 1-5.  At a glance, the Pad\'{e} procedure
dramatically improves the accuracy of the naive perturbation
approximations.  This accuracy is almost the same order as
Zel'dovich-type approximations.

The limit of one-dimensional collapse is achieved by setting $e
\rightarrow 1$ in equation (\ref{eq:m11}) i.e. $h_{\rm in} \rightarrow
-1$.  In this limit, the equation (\ref{eq:mm2}) reduces to the exact
solution, $\delta = \pm a/(1 \mp a)$.  Therefore the Pad\'e
approximation in the present model also has the one-dimensional-exact
property as ZA.  \footnote{ The Pad\'e approximation based on
(\ref{eulerexp2}) also has the one-dimensional-exact property and has
almost the
same behavior as (\ref{eq:mm2}). %
}

\section{Why are Zel'dovich-type approximations so good?}
\setcounter{equation}{0}

In this section, we now discuss the main theme of this paper ``why are
Zel'dovich-type approximations so good?'' based on the preceding two
sections.

As we can observe in Figures 1-5, the curves of approximations are
almost smooth and the accuracy of each approximation can be observed
at any density contrast.  Therefore, we decide to measure the accuracy
of each approximation with each initial axis-ratio by the quantity
$\delta_{\rm approx}/\delta_{\rm true}$ at $\delta_{\rm true}=4$, well
within the nonlinear regime.  This measure of accuracy, in logarithmic
scale, is plotted in Fig. 6, where the horizontal axis is the initial
axis-ratio of spheroidal perturbations.  In the similar way, we plot
the measure of accuracy $\delta_{\rm approx}/\delta_{\rm true}$ at
$\delta_{\rm true}=- 0.6$ for negative density perturbations (voids)
in Fig. 7.

These figures 6 and 7 summarize all of our analysis.  We observe from
them the following facts:
\begin{itemize}
\item[(i)] Zel'dovich-type approximations are far better than any
  Eulerian approximations (except the Pad\'e approximation) in the
  spheroidal model as in the spherical model.  Higher order
  Zel'dovich-type approximations are definitely better than lower
  order Zel'dovich-type approximations all the time.
\item[(ii)] The hierarchy in accuracy of each approximation remains
  the same for all initial axis-ratio of spheroidal positive as well
  as negative perturbations .
\item[(iii)] The accuracy of Zel'dovich-type approximations becomes
  better in both prolate and oblate initial conditions compared with
  the spherical symmetric perturbations.  Moreover the accuracy in
  oblate initial conditions is better than that in prolate initial
  conditions.  All the other Eulerian approximations except Pad\'e
  approximation, on the other hand, have exactly the opposite
  tendency; the accuracy of them becomes worse in both prolate and
  oblate initial conditions.  Moreover the accuracy in oblate initial
  conditions is worse than that in prolate initial conditions.
\item[(iv)] Pad\'e approximations are better than any other Eulerian
  approximations. Higher order Pad\'e approximations are definitely
  better than lower order Pad\'e approximations all the time. Pad\'e
  approximations also have the one-dimensional-exact property.
\item[(v)] The accuracy of Zel'dovich-type approximations and that of
  Pad\'e approximations are almost the same.
\item[(vi)] Zel'dovich-type approximations and the Pad\'e
  approximations approach the exact solution from opposite directions.
  For example in Fig. 6, all the Pad\'e approximations overestimate
  the exact solution while all the Zel'dovich-type approximations
  underestimate the exact solution.  Similarly in Fig. 7, PPZA
  overestimate the exact solution while the corresponding third-order
  Pad\'e approximation underestimate the exact solution.
\end{itemize}

Results (i) and (ii) make us confirm the excellence of the
Zel'dovich-type approximations also in the spheroidal perturbations.
We also confirm the consistency of the higher order Zel'dovich-type
approximations; higher order iteration always yields better accuracy.
Result (iii) supports our first aspect that the validity of the
Zel'dovich-type approximations is grounded on the
one-dimensional-exact property of them.  Actually, the Zel'dovich-type
approximations gradually become much accurate when the system
gradually deviates from the spherical symmetry.  They are most
accurate in oblate collapse, which is effectively dimension one.  They
are second most accurate in prolate collapse, which is effectively
dimension two.  Result (iv) reminds us of the potentiality and the
consistency of the Eulerian scheme.  Result (v) disproves our second
aspect that the validity of Zel'dovich-type approximations is grounded
on their Lagrangian scheme.  Result (vi) signifies that
Zel'dovich-type approximations and Pad\'e approximations are
definitely different scheme despite their similar behavior in Figs. 6
and 7 and their similar appearance in their expansions [(\ref{eq:mm1})
and (\ref{eq:mm2})].

Therefore our results support the aspect (2) that the validity of the
Zel'dovich-type approximations is due to the one-dimensional-exact
property of them and disfavor the aspect (1) that the validity is
grounded on the Lagrangian scheme of them.

\begin{figure}
\begin{center}
   \leavevmode\psfig{figure=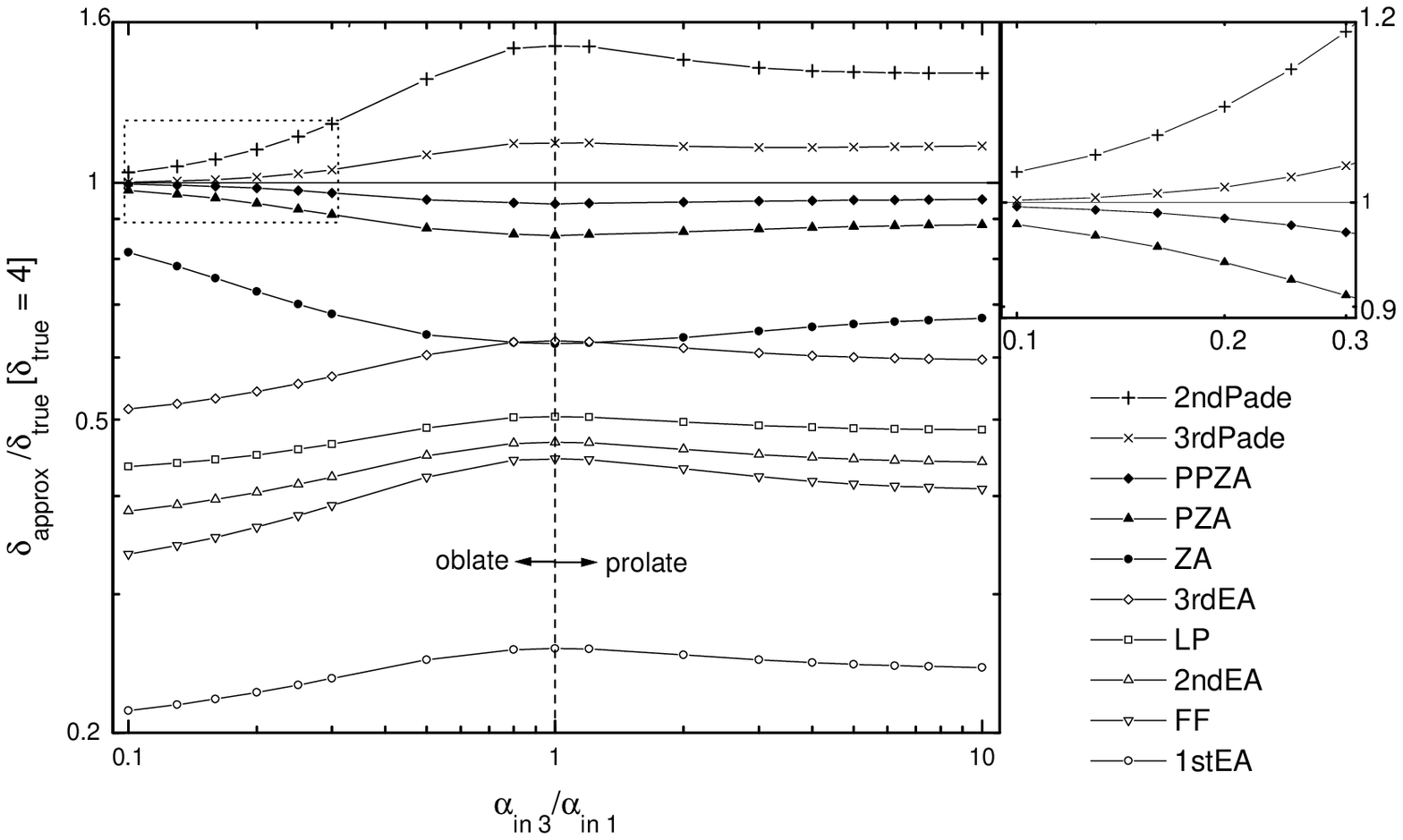,width=15.8cm}
\end{center}
\caption{Density contrast of various approximations in
spheroidal collapse of positive perturbations.  This figure summarizes
all the positive perturbations with various initial axis-ratios.  The
horizontal axis represents the initial axis-ratio $\alpha_{\rm
in3}/\alpha_{\rm in1}$ and the vertical axis represents the ratio of
density contrast $\delta_{\rm approx}/\delta_{\rm true}$ evaluated at
$\delta_{\rm true}=4$.Both the axis ratio and density ratio are in
logarithmic scale in this figure. The right square panel is the
magnification of the part surrounded by the dotted line. \label{fig6}}
\end{figure}
\begin{figure}
\begin{center}
   \leavevmode\psfig{figure=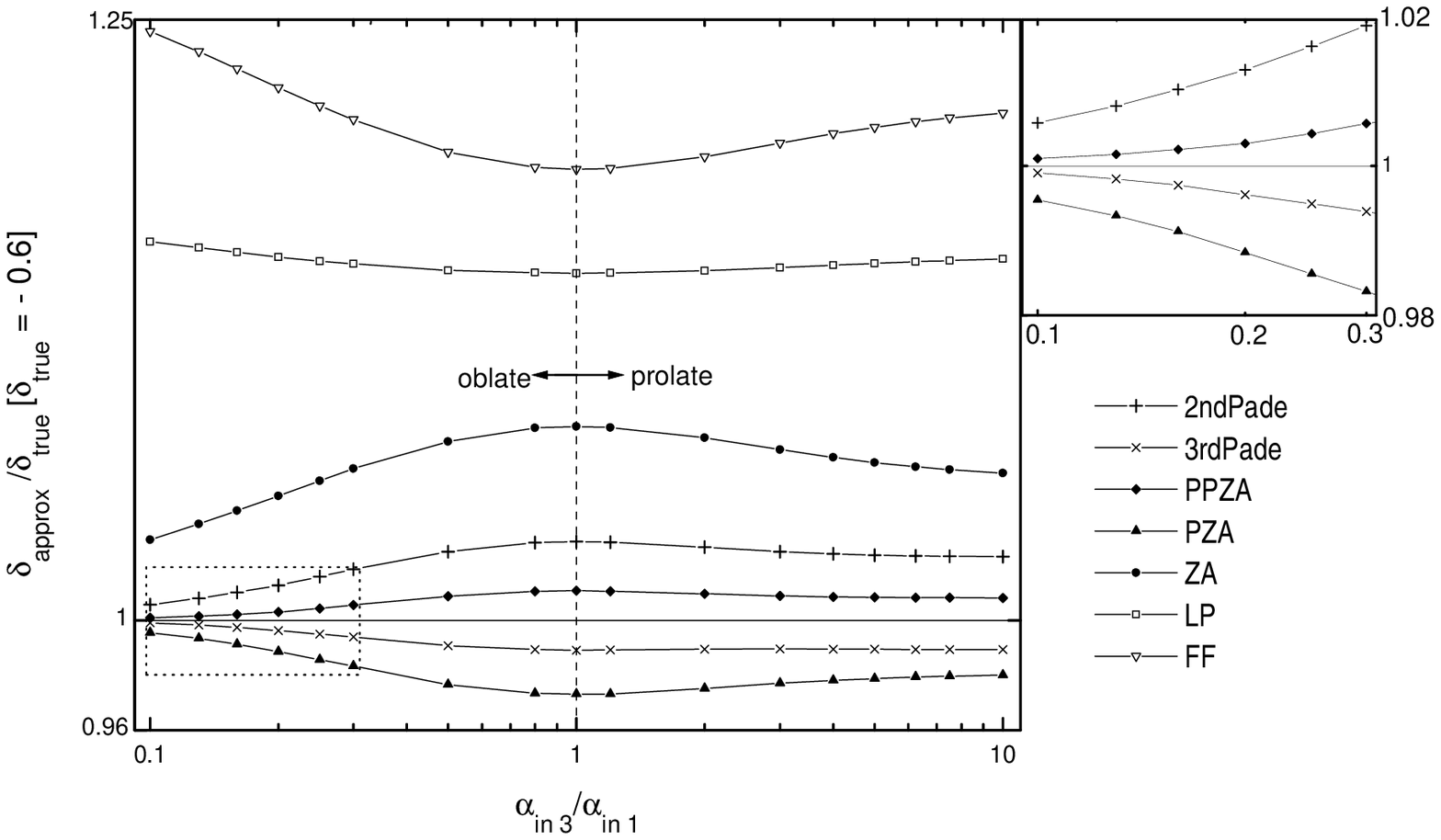,width=15.8cm}
\end{center}
\caption{Density contrast of various approximations in
spheroidal collapse of negative perturbations.  This is the same as
Fig. 6 but for negative perturbations.  The horizontal axis represents
the initial axis-ratio $\alpha_{\rm in3}/\alpha_{\rm in1}$ and the
vertical axis represents the ratio of density contrast $\delta_{\rm
approx}/\delta_{\rm true}$ evaluated at $\delta_{\rm true}=-
0.6$. Both the axis ratio and density ratio are in logarithmic scale
in this figure. The 1st $\sim$ 3rd Eulerian approximations are so bad
that they are outside of this figure. \label{fig7}}
\end{figure}

\section{Conclusions and Discussions}
\setcounter{equation}{0}

In this paper, we have discussed the validity of the Zel'dovich-type
approximations from the two aspects; (1) the dimensionality of the
model and (2) the Lagrangian scheme on which the Zel'dovich-type
approximations grounded.  We introduced a model of spheroidal mass
perturbations and compared several nonlinear approximations as well as
Zel'dovich-type approximations.

We found the following facts from the spheroidal models.  The accuracy
of Zel'dovich-type approximations and Pad\'e approximations becomes
better in both prolate (effective dimension two) and oblate (effective
dimension one) initial conditions compared with the spherical
symmetric perturbations.  Moreover the accuracy in oblate initial
conditions is better than that in prolate initial conditions. These
results may show that the Zel'dovich-type approximations are more
accurate in lower dimensional collapse.  All the other Eulerian
approximations, on the other hand, have exactly the opposite tendency;
the accuracy of them becomes worse in both prolate and oblate initial
conditions.  Moreover the accuracy in oblate initial conditions is
worse than that in prolate initial conditions.

The above facts are consistent with the aspect (1) that the validity
of the Zel'dovich-type approximations is due to the
one-dimensional-exact property of them. On the other hand, the above
facts conflict with the aspect (2) that the validity is grounded on
the Lagrangian scheme of them. For the final confirmations of the
aspect (1), we need further analysis on much wider class of models.

The last fact (vi) in the previous section suggests a possibility to
construct a better approximation beyond Zel'dovich-type approximations
by applying Pad\'e method.  Since Zel'dovich-type and Pad\'e
approximations are definitely different things, we may obtain new
information when the Pad\'e procedure is applied on the
Zel'dovich-type approximations.  The considerations on this
possibility will be affirmatively reported in our separate
publications.

For the actual analysis of observational quantities in the Universe,
it is advantageous to use Eulerian scheme rather than Lagrangian
scheme because the technique of classical field theory is most
naturally applied in the former scheme.  In this context, Pad\'e
approximations in Eulerian scheme, as they are accurate as
Zel'dovich-type approximations, may open new perspective in the
analysis of growth of density fluctuations in the expanding Universe.
Of course we should keep in mind that the validity of the Pad\'e
approximation has not yet been fully established and therefore we
should avoid blind applications of Pad\'e method in cosmology.

\bigskip

We thank Masaaki Morita for helpful discussions on the dimensionality
at the occasion of Tokyo Seminar at Mitaka National Observatory May
1997.
We also thank Oki Nagahara for many suggestions in numerical
calculations.

\newpage

\centerline {\bf REFERENCES}
\bigskip

\def\apjpap#1;#2;#3;#4; {\pp#1, {#2}, {#3}, #4}
\def\apjbook#1;#2;#3;#4; {\pp#1, {#2} (#3: #4)}
\def\apjbok#1;#2;#3;#4;#5; {\pp#1, {#2} (#3; #4: #5)}
\def\apjproc#1;#2;#3;#4;#5;#6; {\pp#1, {#2} #3, (#4: #5), #6}
\def\apjppt#1;#2;#3; {\pp#1, #2, #3}

\baselineskip=17pt

\apjpap Bagla, J. S., \& Padmanabham, T. 1994;MNRAS;266;227;
\apjpap Bernardeau,~F. 1994;ApJ;427;51;
\apjpap Bond, J. R., \& Myers, S. T. 1996;ApJ;103;1;
\apjpap Brainerd,~T.~G., Scherrer,~R.~J., \& Villumsen,~J.~V. 
1993;ApJ;418;570;
\apjbook Binney,~J., \& Tremaine,~S. 1987;Galactic Dynamics;
Princeton, NJ;Princeton Univ. Press;
\apjpap Bouchet,~F.~R., Colombi,~S., Hivon,~E., \& Juszkiewicz,~R. 1995;
A \& A;296;575;
\apjpap Buchert, T. 1992;A \& A;223;9;
\apjpap Buchert, T. 1994;MNRAS;267;811;
\apjpap Buchert, T., \& Ehlers, J. 1993;MNRAS;264;375;
\apjpap Catelan, P. 1995;MNRAS;276;115;
\apjpap Eisenstein,~D.~J., \& Loeb,~A. 1995;ApJ;439;520;
\apjpap Fry, J. N. 1984;ApJ;279;499;
\apjpap Goroff, M. H., Grinstein, B., Rey S.-J., \& Wise M. B. 
1986;ApJ;311;6;
\apjpap Icke, V. 1973;A \& A;27;1;
\apjbook Kellogg,~O. 1953;Foundation of Potential Theory;New York;Dover;
\apjpap Lin, C. C., Mestel, L., \& Shu, F. H. 1965;ApJ;142;1431;
\apjpap Lynden-Bell, D. 1962;Proc. Cambridge Phil. Soc.;50;709;
\apjpap Lynden-Bell, D. 1964;ApJ;139;1195;
\apjpap Matarrese,~S., Lucchin,~F., Moscardini,~L., \& Saez, D. 1992;
MNRAS;259;437;
\apjppt Monaco, P. 1997;MNRAS;in press;
\apjpap Munshi,~D., Sahni,~V., \& Starobinsky,~A.~A. 1994;ApJ;436;517;
\apjbook Peebles,~P.~J.~E. 1980;The Large-Scale Structure of Universe;
Princeton, NJ;Princeton Univ. Press;
\apjbok Press, W. H., Teukolsky, S. A., Vetterling, W. T., \& Flannery, B. 
P.
 1992;Numerical Recipes in Fortran;
2nd ed.;Cambridge;Cambridge Univ. Press;
\apjpap Sahni,~V., \& Coles,~P. 1995;Phys. Rep.;262;1;
\apjpap Sahni,~V., \& Shandarin,~S. 1996;MNRAS;282;641;
\apjpap White, S. D. M., \& Silk, J. 1979;ApJ;231;1;
\apjpap Zel'dovich, Ya. B. 1970;A \& A;5;84;
\apjpap Zel'dovich, Ya. B. 1973;Astrophysics;6;164;

\end{document}